# Auction Based Approach For Resource Allocation In D2D Communication

Saumya Borwankar
Student, Electronics and
Communication Department
Institute of Technology
Nirma University
Ahmedabad, India
17bec095@nirmauni.ac.in

*Abstract*—Device to device communication has prevailed as an issue for small cell networks. Here we have implemented a new scheme that allows us to improve spectral capabilities of mobiles communicating with each other (peer to peer network) for downlink cellular network. Previously the spectral capabilities were handled by Reverse Iterative combinatorial auction mechanism, where the cellular uses used to bid for d2d links. We have made a comparison between Reverse Iterative combinatorial Auction (R-ICA) and New Auction method on the basis of plots on sum rate over SINR and number of d2d users.

## I. Introduction

Small cell utilities are regarded as popular matters to be resolved as well as improved and the effective usage of spectrum for drastically expanding data rate is requested. In a standard small cell network, user equipment interacts with other user equipment through several hops through the base station, even when in proximity, which in turn results in decreased efficiency of the establishment in power utilized and spectral bandwidth. In a D2D established link, two UE's (user equipment) could directly establish a link using cellular resources without the BS (base station) among each other. The BS would then keep a track of the communication's health periodically.

The existent researches grant D2D to increase spectral efficiency if used as an underlay to cellular networks. In D2D communication two UE's communicate without the BS directly by sending data signals using cellular resources, which is quite different from femtocell, in which UE's communicate using cellular base stations that are low powered. The BS controls D2D user communication. Consequently, the capability of getting better spectral utilization has motivated much work in ongoing years, which would eventually lead to improved performance in the system when using D2D that would reuse cellular resources. As an outcome, D2D is expected to be a base component featured in future-generation cellular networks. D2D communication brings benefits to the system capacity as well as improving spectral efficiency while also interfering with the cellular network because of spectrum sharing. Therefore, coordination must be planned to ensure the goal of the improved performance level of cellular communication.

Here in this paper, diverse resource allocation methods for optimum utilization of spectrum has been looked at by the means of a reduction in interference, the complexity of cellular networks, noise. A comparative algorithm is proposed which depends on the auction of spectral bundles by the methods at some definitive price, with an increase in spectral packets for increased price and vice versa. Previous work on D2D communication suggests that with accurate management of resources and interference being restricted between D2D clients and cellular networks, D2D network can increase the throughput of the system. Be that as it may, the issue of allocating resources from cellular users to D2D transmission is complex.

The method used in the paper differs from others in ways to look at sum rate as a priority and ensuring its maximization.

## II. Relevant Work

Work done in MIMO transmission is done in [17] which aims at reducing interference and crosstalk from cellular network downlink to D2D receivers who use the same resources. Ways to improve power efficiency has been looked at in [11, 14, 17] by researchers. Management of interference between D2D users to cellular users and cellular users to D2D users is considered in [12]. For the improvement in gain by spectrum reusing of intra-cell, joining the cellular and D2D users for resource sharing has been looked at in [15]. In [16], the authors have looked at a new alternative greedy heuristic algorithm for reducing interference in cellular networks with the help of channel state information (CSI). Making the uplink (UL) frequency bands efficient with the use of resource allocation that averts interference while also identifying the interfering cellular network, this was studied in [13]. Authors in [17] have provided a study on power management between D2D connection and cellular connections that have common resources that are used for various sharing modes and ideal resource sharing, while also carrying out the performance of the D2D system in both the Manhattan grid environment and a single cell. In [18] the authors describe the findings where multi input multi output cellular network is concerned.

The main idea is to further improve resource management and allocation in users using similar resources. The previous work suggests that D2D communication can vastly increase system throughput by proper resource allocation and management and restricting interference between D2D transmissions and cellular networks.

### A. Reverse Iterative Combinational Auction Mechanism

The D2D communication is a distributive network so the allocation of resources is a major task, auctioning of resources for allocation can improve spectral performance. In [2], the authors proposed R-ICA mechanism for improving spectral efficiency (sum rate) of a D2D network with cellular presence. This auction method, resource units that are spectral resources, which as bidders, fight to get the bundles (D2D pairs) in every auction level. We initialize the value of each resource package followed by a non-monotonic decreasing price-based auction algorithm which is based on a utility function that shows the gain from the channel of the D2D and the cost of the system.

The spectral efficiency is increased as seen from the simulations of the auction process.

On the other hand, the R-ICA algorithm is based on a system where the price paid by the bidder is more important than the sum rate. The sum rate of the system could decrease in spite of the bidder paying a high price. While the bidder paying less price could be able to improve the sum rate of the system. We have introduced a new algorithm, which is simulated in MATLAB and the results are compared with the old auction method.

*B. New Auction Method*

The Reverse Iterative Combinatorial Auctions(R-ICA) mechanism gives an improvement in increasing spectral efficiency, but this method places money given by bidders above the sum rate. This new approach helps cause a decreased sum rate in spite of the bidder paying a high price, while the bidder that pays less price may end up with an increased sum rate which might even be higher than the highest bid price. We propose a new algorithm to overcome this problem.

The new iterative auction method is the algorithm which is based on the calculation of sum rate for the same priced D2D link and allowing the link to the user that causes an increased sum rate, and not depending on the price that an individual user is paying. The auction stays until every link is auctioned. The flow chart in the following figure will show the process.

This new algorithm proposed aims at a different way to allocate resources. The bidders initially bid for the D2D links, what our algorithm does is, it calculates the sum rate for every individual user and looks at the hikes in spectral efficiency while allocating the links to the respective user.

The proposed algorithm is better than the R-ICA algorithm because our algorithm is beneficial to all bidders rather than just the bidder that is paying the highest price for the D2D links. The algorithm is quite similar to R-ICA, here also the bidder's bid in the form of {link number, price}. After which the spectral efficiency is calculated individually if the increase is maximum the link is given to the user even though the price paid by the bidder is not highest. This algorithm promises that every D2D user and cellular network user is benefitted.

III. RESULTS

Simulations have been made in MATLAB software. The graph for sum rate for respective number of D2D users with 4 cellular pair is shown in Fig.1. Averaged SINR statistics of user equipment 1 is shown in Fig.2.

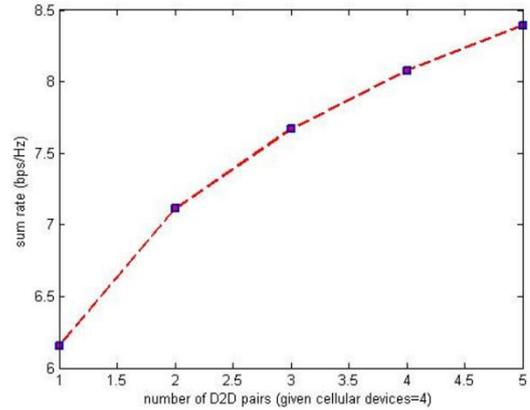

Fig.1 Plot of sum rate vs. number of D2D pairs with 4 cellular clients

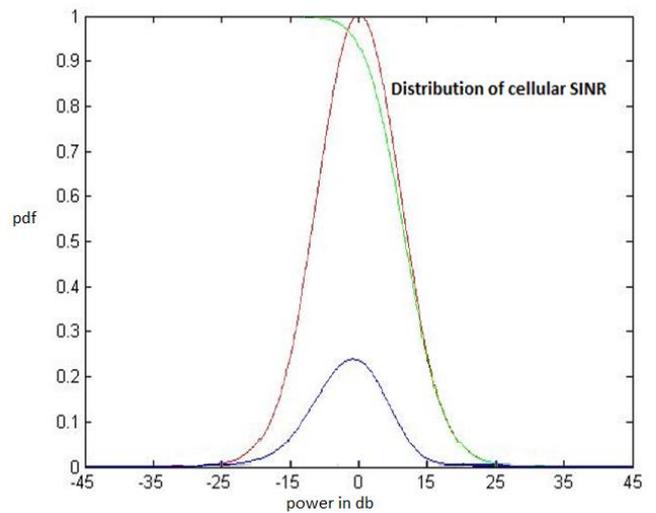

Fig.2 Plot of SINR (averaged) statistics of UE1

A comparison between randomly allocated users and system with 3 and 4 cellular users and their sum rate is plotted in Fig.3. The plot of increasing D2D user with their respective increase sum rate is shown in Fig.4. In the end, the simulation of our new auction method is compared with the previous R-ICA method and the graph is plotted in Fig.5.

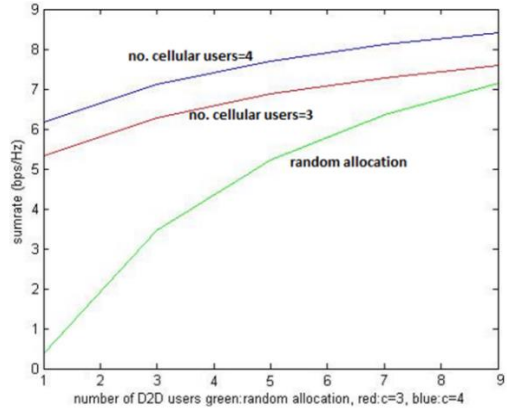

Fig.3 Comparison between randomly allocated users and systems having c=3 and c=4 (R-ICA)

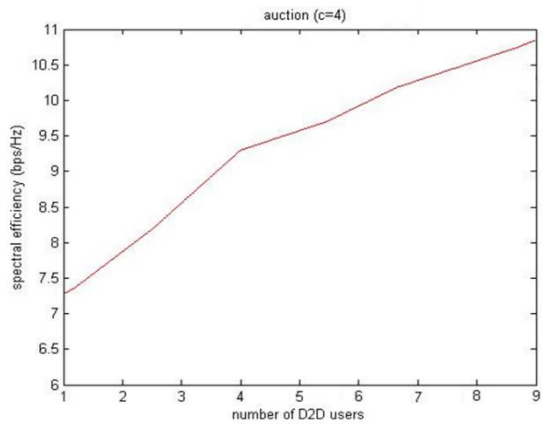

Fig.4 Plot of spectral efficiency vs. number of D2D clients with 4 cellular clients

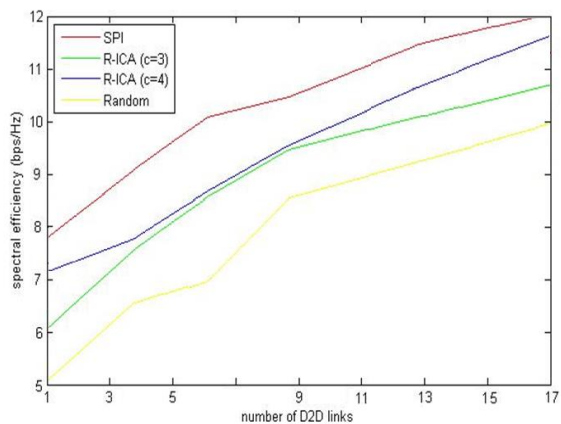

Fig.5 Comparison between newly proposed method and R-ICA method for auction

## IV. CONCLUSION

Device to Device communication aims at providing improved services with the help of already existing cellular networks. The analysis showed that the right arm of the pdf distribution curve is falling steeper than the left arm. This happens due to the log-normally distributed interference. The curve of the D2D clients with various numbers of cellular clients with R-ICA method shows that the R-ICA algorithm increases the sum rate but with increased D2D links the sum rate saturates.

Finally, the new iterative algorithm is proposed which aims at a different approach at the auction process of D2D links which increases the sum rate of the system with the D2D users being benefitted as well as the cellular users as compared to the previous R-ICA algorithm.

REFERENCES

[1] C.-H. Yu, O. Tirkkonen, K. Doppler, and C. Ribeiro, "On the performance of device-to-device underlay communication with simple power control," in Proc. IEEE Vehicular Technology Conference 2009-Spring, Barcelona, Apr. 2009.

[2] Xu, Chen, Lingyang Song, and Zhu Han. "Resource management for device-to-device underlay communication" Berlin: Springer, 2014.

[3] K. Doppler, M. Rinne, C. Wijting, C. Ribeiro, and K. Hugl, "Device-to-device communication as an underlay to LTE-advanced networks," IEEE Commun. Mag., vol. 47, no. 12, pp. 42-49, Dec. 2009.

[4] K J. Zhang, and G. de la Roche, Femtocells: Technologies and Deploymet, John Wiley & Sons Ltd., Nov. 2009.

[5] T. Koskela, S. Hakola, T. Chen, and J. Lehtomaki, "Clustering concept using device-to-device communication in cellular system," in Proc. IEEE Wireless Communications and Networking Conference, Sydney, Apr. 2010.

[6] K. Doppler, M. Rinne, P. Janis, C. Ribeiro, and K. Hugl, "Device-to-device communications; functional prospects for LTE-advanced networks," IEEE International Conference on Communications Workshops, Dresden, Jun. 2009.

[7] S. Hakola, C. Tao, J. Lehtomaki, and T. Koskela, "Device-to-device (D2D) communication in cellular network - performance analysis of optimum and practical communication mode selection," IEEE Wireless Communications and Networking Conference, Sydney, Apr. 2010.

[8] H. Min, W. Seo, J. Lee, S. Park, and D. Hong, "Reliability improvement using receive mode selection in the device-to-device uplink period underlaying cellular networks," IEEE Transactions on Wireless Com- munications, vol. 10, no. 2, pp. 413-418, Feb. 2011.

[9] K. Doppler, C.-H. Yu, C. Ribeiro, and P. Janis, "Mode selection for device-to-device communication underlaying an LTE-advanced network," IEEE Wireless Communications and Networking Conference, Sydney, Apr. 2010.

[10] C.-H. Yu, K. Doppler, C. Ribeiro, and O. Tirkkonen, "Performance impact of fading interference to Device-to-Device communication under- laying cellular networks," IEEE 20th International Symposium on Per- sonal, Indoor and Mobile Radio Communications, pp. 858-862, Tokyo, Sept. 2009.

[11] C.-H. Yu, O. Tirkkonen, K. Doppler, and C. Ribeiro, "Power optimization of device-to-device communication underlaying cellular communi- cation," in Proc. IEEE International Conferemce on Communications, Dresden, Jun. 2009.

[12] T. Peng, Q. Lu, H. Wang, S. Xu, and W. Wang, "Interference avoidance mechanisms in the hybrid cellular and device-to-device systems," IEEE 20th International Symposium on Personal, Indoor and Mobile Radio Communications, pp. 617-621, Tokyo, Sept. 2009.

[13] S. Xu, H. Wang, T. Chen, Q. Huang, and T. Peng, "Effective interference cancellation scheme for device-to-device communication underlay- ing cellular networks," in Proc. IEEE Vehicular Technology Conference 2010-Fall, Ottawa, Sept. 2010.

[14] H. Xing, and S. Hakola, "The investigation of power control schemes for a device-to-device communication integrated into OFDMA cellular system," IEEE 21st International Symposium on Personal Indoor and Mobile Radio Communications, pp. 1775-1780, Instanbul, Sept. 2010.

[15] P. Janis, V. Koivunen, C. Ribeiro, J. Korhonen, K. Doppler, and K. Hugl, "Interference-aware resource allocation for device-to-device radio underlaying cellular networks," in Proc. IEEE Vehicular Technology Conference 2009-Spring, Barcelona, Apr. 2009.

[16] M. Zulhasnine, C. Huang, and A. Srinivasan, "Efficient resource alloca- tion for device-to-device communication underlaying LTE

network," in IEEE 6th International Conference on Wireless and Mobile Computing, Networking and Communications, pp. 368-375, Niagara Falls, Oct. 2010.

[17] P. Janis, V. Koivunen, C.B. Ribeiro, K. Doppler, and K. Hugl, "Interference-avoiding MIMO schemes for device-to-device radio under- laying cellular networks," IEEE 20th International Symposium on Per- sonal, Indoor and Mobile Radio Communications, pp. 2385-2389, Tokyo, Sept. 2009.

[18] Y. Ni, Y. Wang and H. Zhu, "Interference cancellation in device-to-device communications underlying MU-MIMO cellular networks," in China Communications, vol. 16, no. 4, pp. 75-88, April 2019.